\documentclass[a4paper,12pt]{article}
\usepackage[utf8]{inputenc}
\usepackage[T1]{fontenc}
\usepackage{amsmath,amsfonts,amssymb}
\usepackage{color,graphicx}
\usepackage[colorlinks=true,allcolors=black]{hyperref}
\usepackage{bm}
\usepackage{float}
\usepackage{mathrsfs}
\usepackage[font=small,labelfont=bf]{caption}
\usepackage{cite}

\DeclareMathAlphabet{\mathpzc}{OT1}{pzc}{m}{it}

\newcommand{\cS}{\mathcal{S}}
\newcommand{\pR}{\mathcal{R}}
\newcommand{\pa}{\partial} 

\definecolor{pink}{rgb}{1,0,1}

\title{Decisions and disease: a mechanism for the evolution of cooperation}

\author{Carl-Joar Karlsson\textsuperscript{1+}, Julie Rowlett\textsuperscript{1+*}\\
\small \textsuperscript{1}Department of Mathematical Sciences,\\
\small Chalmers University of Technology and The University of Gothenburg\\
\small SE-41296, Gothenburg\\
\small \textsuperscript{*}Corresponding author:  julie.rowlett@chalmers.se\\
\small \textsuperscript{+}The authors contributed equally to this work.}

\begin{document}
\maketitle
\begin{abstract}
	In numerous contexts, individuals may decide whether they take actions to mitigate the spread of disease, or not.  Mitigating the spread of disease requires an individual to change their routine behaviours to benefit others, resulting in a `disease dilemma' similar to the seminal prisoner's dilemma. In the classical prisoner's dilemma,  evolutionary game dynamics predict that all individuals evolve to ‘defect.'  We have discovered that when the rate of cooperation within a population is directly linked to the rate of spread of the disease, cooperation evolves  under certain conditions. For diseases which do not confer immunity to recovered individuals, if the time scale at which individuals receive information is sufficiently rapid compared to the time scale at which the disease spreads, then cooperation emerges.  Moreover, in the limit as mitigation measures become increasingly effective, the disease can be controlled, and the rate of infections tends to zero.  Our model is based on theoretical mathematics and therefore unconstrained to any single context.  For example, the disease spreading model considered here could also be used to describe social and group dynamics.  In this sense, we may have discovered a fundamental and novel mechanism for the evolution of cooperation in a broad sense.   
\end{abstract}

\thispagestyle{empty}

\noindent
\textit{Keywords:} Prisoner's dilemma, evolutionary game dynamics, replicator equation, compartmental models, disease spread, evolution of cooperation

\section*{Introduction}
\begin{figure}
\centering
\includegraphics[width=0.8\textwidth]{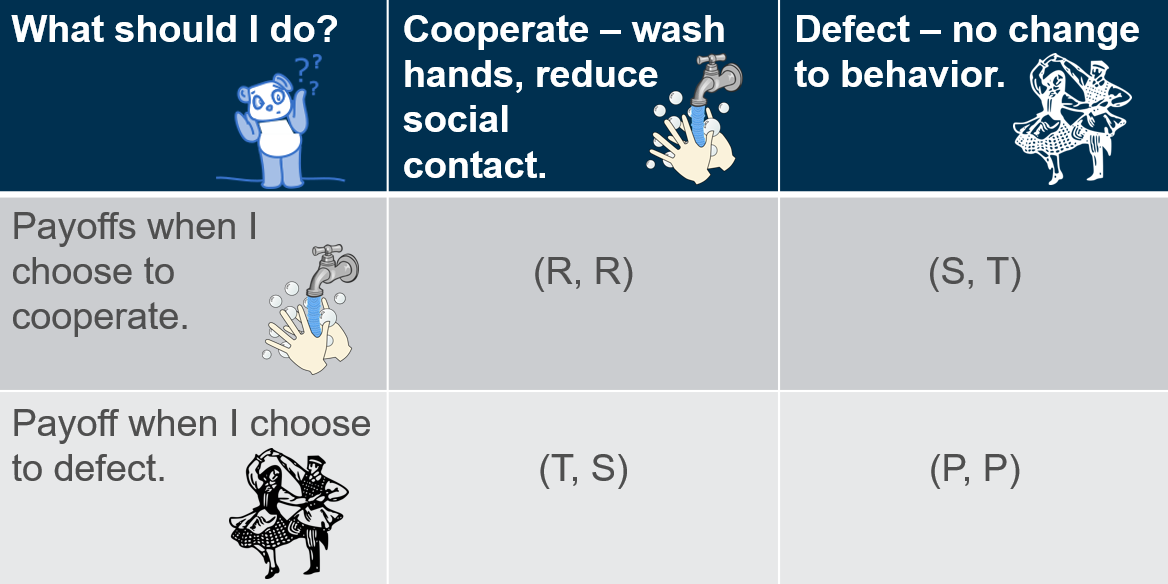}
\caption{In the `disease dilemma' two people have the choice to cooperate, mitigating the spread of the disease, or defect, making no change to their regular behaviour.  This is described by the two player non-cooperative game shown here in normal form.  Image source and license:  \href{https://openclipart.org/}{\texttt{openclipart.org}}, \href{https://creativecommons.org/publicdomain/zero/1.0/deed.en}{CC0 1.0}.}
\label{fig:dd}
\end{figure}

Decisions made by individuals affect the population, not the least in disease spreading. Several researchers have investigated the interplay between diseases and decisions by combining compartmental models with game theory \cite{funk2010modelling,verelst2016behavioural,weston2018infection,reluga2010,chen2011public}.  Common considerations are dynamics on networks or lattices \cite{rhodes1997epidemic,zhao2015,XIA2019185,gomez2010discrete,kan2017effects,hota2019,zhang2014effects,schimit2011vaccination,trajanovski2015decentralized,chen2011public,liu2012impact,kabir2019,kuga2019vaccinate,fu2011imitation} and well-mixed populations \cite{hayashi2016effects,bhattacharyya2011wait,bauch2012evolutionary,bauch2005imitation,li2017provisioning,wang2013impact,wang2020effect,jnawali2016emergence,wu2011imperfect,cai2014effect}. The former's strength is that it captures the effect of population structures, while the latter's strength is that it highlights the individuals' perception of the payoff. We consider a well-mixed society in which individuals choose to what extent they will exert preventive measures to mitigate the spread of an infectious disease.  There are two choices:  exert mitigating measures to prevent the spread of the disease, and do-not-mitigate, making no efforts to prevent the spread of the disease.  The World Health Organisation \cite{who} and numerous other references including \cite{ebola,asiatimes,stubborn} argue that it is reasonable to describe this situation with the Prisoner's Dilemma (PD); this is depicted in Figure~\ref{fig:dd}.  

The payoffs may be interpreted in the sense that if both Alice and Bob cooperate, then they are behaving towards their mutual common good, but they are also making personal sacrifices by changing their usual behaviour to mitigate disease spread. On the other hand, if Alice cooperates while Bob defects, then Alice’s mitigation efforts prevent Bob from catching the disease, thus Bob benefits from Alices’s cooperation. Alice receives no such benefit from Bob’s reckless actions, therefore Alice is not only restricting her personal freedoms, she is also at risk of catching the disease due to Bob’s recklessness. If both Alice and Bob defect, then they are both at risk of catching the disease, but neither makes any personal sacrifices. Consequently, this situation is described by payoffs which satisfy 
\begin{equation} 
\label{pdpayoffs} 
S<P<R<T.
\end{equation} 

The unique equilibrium strategy of this game is mutual defection, and when this game is used to predict behaviours according to evolutionary game dynamics, the result is always defection \cite{nowak}.  Nonetheless, in many contexts which fit into a PD type game, cooperation may in fact be observed \cite{axelrod,axelrod2,cooperate1,cooperate2,cooperate3,cooperate4,cooperate5}.  In the particular case of the PD, there have been numerous mechanisms proposed for the evolution of cooperation \cite{martin,ore}. To our knowledge, it has been unknown -- until now -- whether cooperation emerges when the payoff is a trade-off between the PD and the effect on disease spreading (changes to the infection transmission rate).
\section*{Methods}
Infections like those from the common cold, flu, and many sexually transmitted infections do not confer any long-lasting immunity, and individuals become susceptible once they recover from infection.  These diseases are described by the SIS compartmental model.  Poletti et al. \cite{poletti2009} implemented a hybrid model in which human decisions affect the rate at which the disease spreads.  They assigned two different rates of infection corresponding to individuals either changing their behaviour to mitigate the spread of the disease, or not doing so.  We follow this approach by assigning the rates of infection for cooperators and defectors, $\beta_C < \beta_D$, respectively.  The infection-producing contacts (via, e.g., droplets from someone who sneezes) per unit time is weighted by the proportion of cooperators, $x$, and defectors, $1-x$, and is therefore 

\begin{equation} 
\beta= (1-x)\beta_D+x\beta_C.
\end{equation} 
We consider three timescales: (i) the disease transmission timescale $t$,  (ii) the timescale at which individuals decide whether or not to mitigate the spread of the disease $\alpha_1 t$, and (iii)  the timescale at which individuals receive the PD payoffs $\alpha_2 t$.  
The replicator equations for our hybrid SIS-PD model are therefore 

\begin{align} 
\frac{dI}{dt}=&\  ((1-x(t))\beta_D+x(t)\beta_C)I(t)(1-I(t))-\gamma I(t), \\
\frac{dx}{dt}=&\ x(t)(1-x(t))\left[\alpha_1(\beta_D-\beta_C)I(t)-\alpha_2 ([T-R]x(t)+[P-S](1-x(t))\right].
\end{align}
Above, the quantities on the left side are differentiated with respect to $t=$ time, $I$ is the portion of the population which is infected, $x$ is the portion of the population which chooses to cooperate, and $\gamma$ is the transmission rate.  If $D$ is the duration of the infection, then $\gamma = 1/D$.  We note that $1-I$ is the portion of the population which is susceptible to infection, since in this model there is no immunity. 

For the sake of simplicity, and since no generality is lost, we shall assume the PD payoffs~(\ref{pdpayoffs}) satisfy 
\(T-R = P-S,\)
leading to the simplified system 

\begin{align} 
\frac{dI}{dt}&= ((1-x(t))\beta_D+x(t)\beta_C)I(t)(1-I(t))-\gamma I(t), \\
\frac{dx}{dt}&=x(t)(1-x(t))  [\alpha_1(\beta_D-\beta_C)I(t)-\alpha_2 (T-R)].  
\end{align} 

Since $\beta_D > \beta_C$, and $T>R$, the terms in the equation for the evolution of cooperators have opposite signs, resulting in a competition between avoidance of  disease carriers and PD reward.  Moreover, the relationship between the decision-making and PD-payoff time scales  is key.  

Similar calculations lead to the replicator equations for the SIR-PD model

\begin{align} \label{sirpd-eqns} 
\dot \cS(t) &= - \left( (1-x(t)) \beta_D + x(t) \beta_C \right) \cS(t) I(t) \\ 
\dot I (t)&=  \left( (1-x(t)) \beta_D + x(t) \beta_C \right) \cS(t) I(t) - \gamma I(t) \nonumber \\
\dot \pR(t) &= \gamma I(t) \nonumber \\ 
\dot x(t) &= x(t)(1-x(t)) \left( \alpha_1(\beta_D - \beta_C) I(t) - \alpha_2(T-R) \right) \nonumber 
\end{align} 
Above, $\pR$ is the number of recovered, deceased, or immune individuals. This model is reasonably predictive for infectious diseases that are transmitted from human to human, and where recovery confers lasting resistance. Since $\dot \cS+\dot I+\dot \pR = 0$, the triplet $(x,I,\cS)$ describes the complete system.

\section*{Results}
For the SIR-PD model, the equilibrium points consist of the set $(x, I, \cS)$ 
\[\left\{ (0, 0, \cS^*), \, (1, 0, \cS^*) \right\} \quad \cS^* \in [0,1].\]
An equilibrium point with $x=0$ is stable if $\beta_D \cS^* \leq \gamma$, and it is unstable if the reverse inequality holds.  All equilibrium points with $x=1$ are unstable.    

\begin{table}[ht]
\centering
\begin{tabular}{|l|l|}
\hline
\textbf{Range} & \textbf{Equilibrium}\\
\hline
$0<\alpha_1<\check{\alpha}_1$ & $(0,1-\gamma/\beta_D)$ \\
\hline
$\check{\alpha}_1\leq\alpha_1\leq\hat{\alpha}_1$ & $(x^*,I^*)$ \\
\hline
$\hat{\alpha}_1<\alpha_1$ & $(1,1-\gamma/\beta_C)$\\
\hline
\end{tabular}
\caption{\label{tab:results}The asymptotically stable equilibrium points of the SIS-PD model in the specified ranges of $\alpha_1$, where $\check{\alpha}_1$ is defined in \eqref{alpha1check}.}
\end{table}

The equilibrium points of the SIS-PD system are the set of $(x, I)$: 
\[\left \{ (0,0),  (1, 0),   \left(0, 1 - \frac \gamma \beta_D \right),  \left(1, 1-\frac \gamma \beta_C \right), \left(x^*, I^*\right) \right\},
\]
where
\[ 
x^* = \frac{\beta_D}{\beta_D - \beta_C} - \frac{\gamma}{(\beta_D - \beta_C)(1-I^*)}, \quad I^* = \frac{\alpha_2(T-R)}{\alpha_1(\beta_D - \beta_C)}.
\]
The equilibrium point, $(x^*, I^*)$, is well-defined as long as $x^* \in [0,1]$, and $I^* \in [0,1),$  since $T>R$, and $\alpha_1, \alpha_2 > 0$. We compute that 
\[ I^* < 1 \iff \frac{\alpha_2 (T-R)}{\beta_D - \beta_C} < \alpha_1.\]
We further compute 
\begin{equation}\label{alpha1check} 0 \leq x^*\leq 1 \iff \check{\alpha}_1 \leq \alpha_1\leq \hat{\alpha}_1, 
\end{equation}
where
\begin{equation}
\check{\alpha}_1= \frac{\beta_D}{\beta_D - \gamma} \frac{\alpha_2 (T-R)}{\beta_D - \beta_C}\quad \mathrm{and}\quad \hat{\alpha}_1= \frac{\beta_C}{\beta_C - \gamma} \frac{\alpha_2 (T-R)}{\beta_D - \beta_C}.
\end{equation}
Since $1<\beta_D/(\beta_D-\gamma)$, this condition immediately implies $I^*<1$.  We note that 
\[ \frac{\beta_D}{\beta_D - \gamma} < \frac{\beta_C}{\beta_C - \gamma} \implies \frac{\beta_D}{\beta_D - \gamma} \frac{\alpha_2 (T-R)}{\beta_D - \beta_C} < \frac{\beta_C}{\beta_C - \gamma} \frac{\alpha_2 (T-R)}{\beta_D - \beta_C}. \]
Whenever it exists, the equilibrium point $(x^*, I^*)$ is \em always \em stable (and asymptotically stable).

The equilibrium point $(0,0)$ is stable (and asymptotically stable) if 
\[ \beta_D < \gamma, \quad T>R.\]
Since $dI/dt<0$ when this condition is satisfied, there is no epidemic.  The equilibrium point $(1,0)$ is never stable for PD payoffs \eqref{pdpayoffs}.  The equilibrium point $(0,1- \gamma/\beta_D)$, is well defined if $\beta_D\geq\gamma$, because $0 \leq I \leq 1$, and it is stable (and asymptotically stable) if 
\[ \alpha_1(\beta_D - \beta_C)(1-\gamma/\beta_D) < \alpha_2 (T-R). \] 
For PD payoffs \eqref{pdpayoffs}, this is equivalent to 
\begin{equation} 
\label{notawesome}  
\alpha_1< \hat{\alpha}_1 
\end{equation}
The equilibrium point $(1,1-\gamma/\beta_C)$ is well defined if $\beta_C\geq\gamma$.  It is stable (and asymptotically stable) if 
\[  \alpha_2 (T-R) < \alpha_1(\beta_D - \beta_C) (1-\gamma/\beta_C).\]
For PD payoffs \eqref{pdpayoffs}, this equilibrium point is stable (and asymptotically stable) when 
\begin{equation} 
\label{awesome}  
\check{\alpha}_1 < \alpha_1. 
\end{equation}

\begin{figure}
	\centering
	\includegraphics[width=0.82\textwidth]{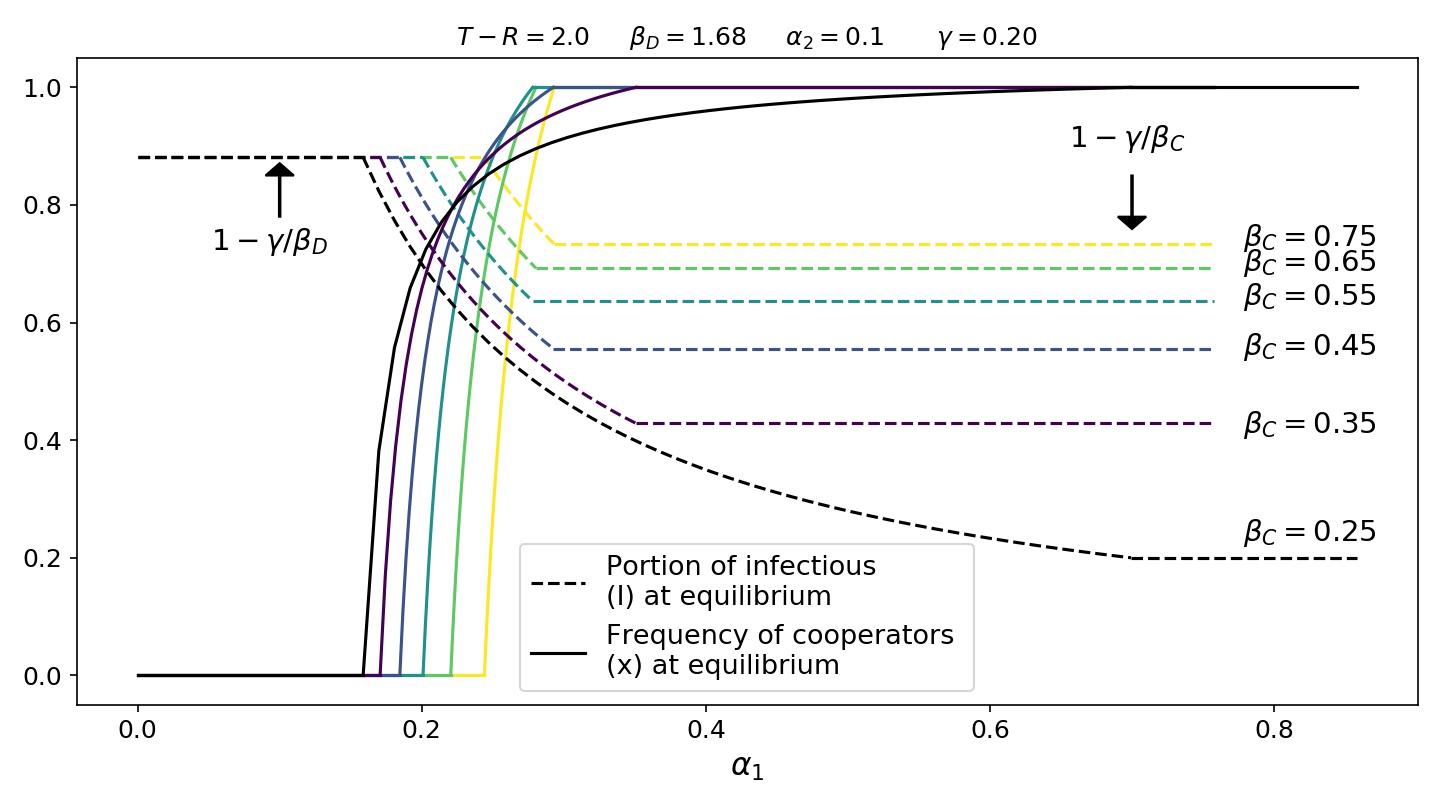}
	\caption{The values of $\beta_D$ and $\gamma$ above were suggested \cite{lin2020conceptual}; however these values can be modified to any disease parameters. Since it is the relationship between $\alpha_1$ and $\alpha_2$, rather than their individual values which affects the dynamics, we simply fix $\alpha_2=0.1$.  The value of $\alpha_1$ ranges along the horizontal axis.  The vertical axis is the frequency within the population.  For sufficiently large $\alpha_1$, the population evolves to cooperation.  At the same time, the more effective the mitigation measures are, the lower $\beta_C$ is, which pushes the portion of infected individuals to zero.  More precisely, when $\alpha_1\geq \hat{\alpha}_1$, then $\lim_{\beta_C \searrow \gamma}I=0$.}
	\label{fig2} 
\end{figure}

\begin{figure}
	\centering
	\includegraphics[width=\textwidth]{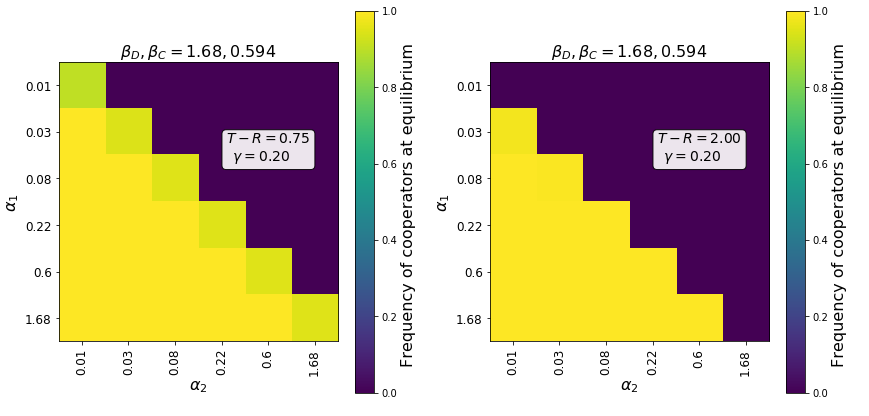}
	\caption{The evolution to cooperation depends on the relationship between  $\alpha_1$ and $\alpha_2$, when the parameters $\beta_D$ and $\gamma$ are as suggested in \cite{lin2020conceptual} and with $T-R=0.75$ (left figure) or $T-R=2$ (right figure).  Here the value of $\beta_C$ corresponds to moderately effective mitigation measures.}
	\label{fig3} 
\end{figure}

\begin{figure}
	\centering
	\includegraphics[width=0.8\textwidth]{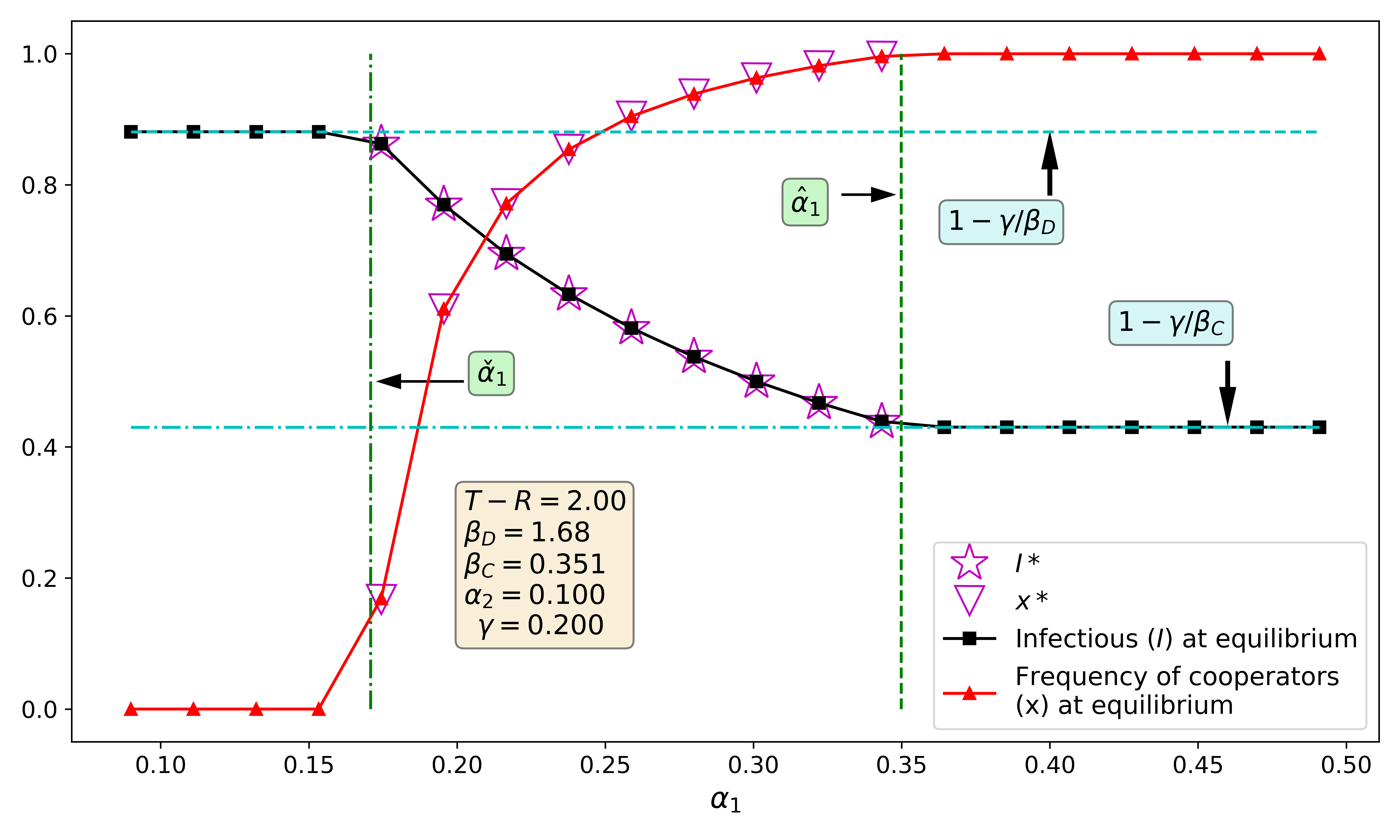}
	\caption{The results from numerical integration agree with the analytical results. The values of $\beta_D$ and $\gamma$ above were suggested \cite{lin2020conceptual}.  Here the value of $\beta_C$ corresponds to mitigation measures which are more effective than in Figure \ref{fig3} but still imperfect.}
	\label{fig:computer-results}
\end{figure}

Our results are summarised in Table~\ref{tab:results}. Figure \ref{fig2} shows how the evolution of cooperation and the rate of infections depend on $\alpha_1$ and $\beta_C$ when $\beta_D=1.68$ and $\gamma=1/5$ as suggested in \cite{lin2020conceptual}.  We note that these values were selected merely for the sake of visualisation, as our theoretical results hold for any parameter values. If both $\alpha_1$ and $\alpha_2$ vary, we obtain convergence to cooperation as shown in Figure \ref{fig3}.  Figure~\ref{fig:computer-results} shows that the numerical integration agrees perfectly with the analytical results.

\section*{Discussion}
It has been suggested that mass media could be used to reduce HIV-infections \cite{hiv}, which fits well with our theoretical model. 
If an infectious disease does not confer immunity to those who recover from it, then SIS is a suitable model. The rate of spread for those who make no mitigation efforts, $\beta_D$, is strictly larger than the rate of spread for those who make mitigation efforts, $\beta_C$.   Our results show that the relationship between the time scale of decision making, $\alpha_1 t$, and the timescale of PD payoffs, $\alpha_2 t$ is crucial.  Decision-making is influenced by the speed at which individuals access or receive information upon which to base their decisions.  It is reasonable to assume that the timescale of PD payoffs is similar to the timescale $t$ for the spread of disease, or at least on the same order of magnitude.  On the other hand, the speed at which individuals can access information could be much faster.  This corresponds to
\( 
\alpha_1 \gg\alpha_2.
\)
When $\alpha_1>\hat{\alpha}_1$, the equilibrium point $(1, 1-\gamma/\beta_C)$ exists.  Consequently, for sufficiently large $\alpha_1$, the \em  unique \em equilibrium point of the system corresponds to total cooperation.  Moreover, in this case the portion of the population which is infected tends to $1-\gamma/\beta_C$.  We therefore also have 
\begin{equation} \label{eq:betaclimit} \lim_{\beta_C \searrow \gamma} 1-\frac{\gamma}{\beta_C} = 0. \end{equation} 
This shows that in the limit towards effective mitigation measures, the rate of the population which is infected tends to zero. We summarise these insights below. 

\begin{quote} \em  In the context of a communicable disease which does not confer immunity, if information is made available to all individuals quickly relative to the spread of the disease, all rational individuals acting in their best self interest will evolve to cooperate. Moreover, if their mitigation efforts are effective, they drive the rate of spread of the disease to zero. \em 
\end{quote} 

These insights suggest a general strategy for controlling both new diseases as well as diseases which are known not to confer immunity.  For a new disease, it is unknown and unknowable whether contracting and recovering from the disease grants immunity \cite{mice}. Moreover, vaccines require time for development and testing \cite{vaccine}. It may therefore be prudent to use the SIS model for new communicable diseases.  Moreover, our results show that the evolution of cooperation does not occur in the SIR-PD model.  Consequently, if the desired outcome is the evolution of cooperation and control of disease, the SIS-PD model yields the best results.  The value of $\alpha_1$ may be associated to the frequency of public service announcements (PSAs) explaining the recommended measures.  The more frequent the PSAs, the higher the value of $\alpha_1$.  Our results prove that when $\alpha_1$ becomes very large, cooperation emerges, and the amount of infections can be controlled.  Moreover, when mitigation measures are made increasingly effective, in the limit the frequency of infections tends to zero.

The perceived benefit of defecting is defined by the PD payoffs~(\ref{pdpayoffs}), so that defecting is still perceived to offer benefits if others cooperate.  The key to the evolution for cooperation is the time scale for decision making.  This can be much faster than the time scale at which one can actually reap the benefits of defecting.  When this is the case, the population evolves towards cooperation.   
Our results are not constrained to any specific disease, but rather suggest a general strategy to promote the evolution of cooperation in the classical Prisoner's Dilemma when linked to the spread of disease according to the SIS model.  The SIS model has further applications to describing social and group dynamics \cite{wang2020effect}.  Our model may thereby provide a mechanism for the evolution of cooperation in social contexts as well.


\section*{Acknowledgements}
We are grateful to Philip Gerlee and Alessandro Bravetti for insightful discussions and suggesting relevant references. We thank Johan Runeson for the idea to use game theory to model human behaviour in the corona pandemic.  Both authors are supported by Swedish Research Council Grant GAAME 2018-03873.


\section*{Supplementary information}
Our equations in the SIS-PD model are:
\begin{equation} 
\frac{dI}{dt}= ((1-x(t))\beta_D+x(t)\beta_C)I(t)(1-I(t))-\gamma I(t),
\end{equation}
\[
\frac{dx}{dt}=x(t)(1-x(t))[\alpha_1(\beta_D-\beta_C)I(t)-\alpha_2 (T-R)].  
\]

\subsection*{Calculation and classification of equilibrium points in the SIS-PD model.} 
We compute that the equilibrium points of the system are the set of $(x, I)$:
\[\left \{ (0,0),  (1, 0),   \left(0, 1 - \frac \gamma \beta_D \right),  \left(1, 1-\frac \gamma \beta_C \right), \left(x^*, I^*\right) \right\} 
\] 
where 
\[ 
x^* = \frac{\beta_D}{\beta_D - \beta_C} - \frac{\gamma}{(\beta_D - \beta_C)(1-I^*)}, \quad I^* = \frac{\alpha_2(T-R)}{\alpha_1(\beta_D - \beta_C)}.
\]

To determine the nature of the equilibrium points, that is whether they are (asymptotically) stable or unstable, we compute the Jacobian matrix
\[ \begin{bmatrix} \frac{\partial f}{\partial I}, & \frac{\partial f}{\partial x} \\ \frac{\partial g}{\partial I}, & \frac{\partial g}{\partial x} \end{bmatrix}
\]
whose entries are
\[
\frac{\partial f}{\partial I} = (1-2I)((1-x)\beta_D + x \beta_C) - \gamma,\]
\[ \frac{\partial f}{\partial x} = I(1-I)(\beta_C - \beta_D) \]
\[
\frac{\partial g}{\partial I} = \alpha_1 x(1-x)(\beta_D - \beta_C),\]
\[ \frac{\partial g}{\partial x}= (1-2x)\left[\alpha_1(\beta_D - \beta_C)I  - \alpha_2(T-R) \right].
\]

At the equilibrium point $(0,0)$ the Jacobian matrix is 
\[
\begin{bmatrix} \beta_D - \gamma & 0 \\ 0 & -\alpha_2(T-R) \end{bmatrix}. 
\] 
If the real parts of all eigenvalues are negative, then the equilibrium point is stable and asymptotically stable.  This holds when 
\[ \beta_D < \gamma, \quad T>R.\]

The Jacobian matrix at the equilibrium point $(1, 0)$ is 
\[
\begin{bmatrix} \beta_C - \gamma & 0 \\ 0 & \alpha_2(T-R) \end{bmatrix}. 
\] 
Since $\alpha_2 > 0$, and $T>R$, this equilibrium point is unstable.  

The equilibrium point $(0, 1-\gamma/\beta_D)$ has Jacobian matrix 
\[
\begin{bmatrix} \gamma - \beta_D & (1-\gamma/\beta_D)(\gamma/\beta_D) (\beta_C - \beta_D) \\ 0 & \alpha_1(\beta_D - \beta_C)(1-\gamma/\beta_D) - \alpha_2 (T-R) \end{bmatrix}\]
Since $0 \leq I \leq 1$, this is well-defined if and only if 
\[ \gamma \leq \beta_D. \] 
It is stable and asymptotically stable if 
\[ \alpha_1(\beta_D - \beta_C)(1-\gamma/\beta_D) < \alpha_2 (T-R). \] 
For PD payoffs, this is equivalent to 
\[ \alpha_1 < \frac{\beta_D - \gamma}{\beta_D} \alpha_2 \frac{T-R}{\beta_D - \beta_C}. \] 

The equilibrium point $(1, 1-\gamma/\beta_C)$ is well-defined if and only if
\[\gamma \leq \beta_C.\]
It has Jacobian matrix 
\[ 
\begin{bmatrix} \gamma - \beta_C & (1-\gamma/\beta_C)(\gamma/\beta_C) (\beta_C - \beta_D) \\ 0 & \alpha_2 (T-R) - \alpha_1(\beta_D - \beta_C) (1-\gamma/\beta_C)
\end{bmatrix}\]

It is stable and asymptotically stable if 
\[  \alpha_2 (T-R) < \alpha_1(\beta_D - \beta_C) (1-\gamma/\beta_C).\]
Note that if one changes the PD payoffs so that $T<R$, then since $\beta_D > \beta_C$, and $\alpha_1, \alpha_2 > 0$, this equilibrium point is always stable.  The condition above is equivalent to 
\[ \frac{\beta_C}{\beta_C - \gamma} \frac{\alpha_2 (T-R)}{\beta_D - \beta_C} < \alpha_1.\] 

The equilibrium point, $(x^*, I^*)$, exists as long as $x^* \in [0,1]$, and $I^* \in [0,1),$  since $T>R$, and $\alpha_1, \alpha_2 > 0$. 
We compute that 
\[ I^* < 1 \iff \frac{\alpha_2 (T-R)}{\beta_D - \beta_C} < \alpha_1.\]
We further compute 
\[ 0 \leq x^* \iff \left( \frac{\beta_D}{\beta_D - \gamma} \right) \frac{\alpha_2 (T-R)}{\beta_D - \beta_C} \leq \alpha_1.\]
Since $1 <  \frac{\beta_D}{\beta_D - \gamma}$, this condition immediately implies $I^* < 1$.  We note that 
\[ \frac{\beta_D}{\beta_D - \gamma} < \frac{\beta_C}{\beta_C - \gamma}\]
\[
 \implies  \left( \frac{\beta_D}{\beta_D - \gamma} \right) \frac{\alpha_2 (T-R)}{\beta_D - \beta_C} < \frac{\beta_C}{\beta_C - \gamma} \frac{\alpha_2 (T-R)}{\beta_D - \beta_C}. \]

We compute the Jacobian matrix 
\[
\begin{bmatrix} - \frac{I^* \gamma}{1-I^*} & I^*(1-I^*)(\beta_C - \beta_D) \\ \alpha_1 x^*(1-x^*)(\beta_D - \beta_C) & 0 \end{bmatrix}, 
\]
Under these conditions, we compute that it is always stable (and asymptotically stable), because we compute that the eigenvalues of the Jacobian matrix have negative real part, since the matrix is of the form 
\[ \begin{bmatrix} - & - \\ + & 0 \end{bmatrix}. \] 

Hence, the interesting values of $\alpha_1$ are 
\[ \alpha_1 <  \left( \frac{\beta_D}{\beta_D - \gamma} \right)  \frac{\alpha_2 ( T-R)}{\beta_D - \beta_C} \implies \exists \left( 0, 1-\frac{\gamma}{\beta_D} \right),\]
and this equilibrium point is stable and asymptotically stable.  When $\alpha_1$ is greater than or equal to this value, 
\[ \left( \frac{\beta_D}{\beta_D - \gamma} \right) \frac{\alpha_2 (T-R)}{\beta_D - \beta_C} \leq \alpha_1 \]
\[
\implies \exists (x^*, I^*) \textrm{ until } \alpha_1 =  \frac{\beta_C}{\beta_C - \gamma} \frac{\alpha_2 (T-R)}{\beta_D - \beta_C},\]
and this equilibrium point is stable and asymptotically stable.  For $\alpha_1 >  \frac{\beta_C}{\beta_C - \gamma} \frac{\alpha_2 (T-R)}{\beta_D - \beta_C}$, $x^* > 1$, so this equilibrium point ceases to exist, but 
\[  \frac{\beta_C}{\beta_C - \gamma} \frac{\alpha_2 (T-R)}{\beta_D - \beta_C} < \alpha_1 \implies \exists \left( 1, 1-\frac{\gamma}{\beta_C} \right),\] 
and this equilibrium point is stable and asymptotically stable.  

\subsection*{Calculation and classification of all equilibrium points in the SIR-PD model}  \label{si2} 
For the SIR-PD model, the equations are
\begin{align} \label{sir-eq-app} \dot \cS &= - ((1-x) \beta_D + x \beta_C)\cS I = \phi(x, I, \cS)\\ 
\dot I &= - \dot \cS - \gamma I = \psi(x, I, \cS), \nonumber \\
\dot \pR &= \gamma I, \nonumber \\ 
\dot x(t) &= x(1-x)\left[ \alpha_1(\beta_D - \beta_C)I - \alpha_2 (T-R) \right] = g(x, I). \nonumber
\end{align} 
The set of equilibrium points for $(x, I, \cS)$ is the set 
\[
\left\{ (0, 0, \cS^*), \, (1, 0, \cS^*) \right\} \quad \cS^* \in [0,1]. 
\]
The Jacobian matrix is 
\[
\begin{bmatrix} \frac{\pa \phi}{\pa \cS}, & \frac{\pa \phi}{\pa I}, & \frac{\pa \phi}{\pa x} \\ \frac{\pa \psi}{\pa \cS},  & \frac{\pa \psi}{\pa I}, & \frac{\pa \psi}{\pa x} \\ 
\frac{\pa g}{\pa \cS}, & \frac{\pa g}{\pa I}, & \frac{\pa g}{\pa x} \end{bmatrix},
\]
and its entries are 
\begin{align*} 
&\frac{\pa \phi}{\pa \cS} = -\left( (1-x) \beta_D + x \beta_C \right) I, \quad \frac{\pa \phi}{\pa I} =  -\left( (1-x) \beta_D + x \beta_C \right) \cS, \quad \frac{\pa \phi}{\pa x} = (\beta_D - \beta_C) \cS I \\ 
&  \frac{\pa \psi}{\pa \cS} =   \left( (1-x) \beta_D + x \beta_C \right) I, 
\quad \frac{\pa \psi}{\pa \cS}  = \left( (1-x) \beta_D + x \beta_C \right)\cS - \gamma, 
\quad \frac{\pa \psi}{\pa x}   = (\beta_C - \beta_D) \cS I \\ 
&\frac{\pa g}{\pa \cS} = 0, \, \frac{\pa g}{\pa I} = x(1-x) \alpha_1(\beta_D - \beta_C), \quad \frac{\pa g}{\pa x} = (1-2x) \left[ \alpha_1(\beta_D - \beta_C)I - \alpha_2 (T-R) \right].
\end{align*} 

The equilibria with $x=0$ have Jacobian matrix 
\[ 
\begin{bmatrix} 0 & -\beta_D \cS^* & 0 \\ 0 & \beta_D \cS^* - \gamma & 0 \\ 0 & 0 & - \alpha_2 (T-R) \end{bmatrix} 
\] 
The equilibrium point is stable and asymptotically stable if $\beta_D \cS^* < \gamma$.  
The equilibria with $x=1$ have Jacobian matrix 
\[ 
\begin{bmatrix} 0 & -\beta_C \cS^* & 0 \\ 0 & \beta_C \cS^* - \gamma & 0 \\ 0 & 0 &  \alpha_2 (T-R) \end{bmatrix} 
\] 
Since $\alpha_2 > 0$, and $T>R$, this equilibrium is always unstable.  

\subsection*{Computer simulation details}
All simulations were performed with the initial conditions $x(0)=0.5$ and $I(0)=0.001$ if otherwise is not stated. The choice of initial conditions does not change the convergence results, only the dynamics at small times. The ODEs are integrated using the Python routine \verb|scipy.integrate.solve_ivp| with default settings. These simulations were used to produce Figure~\ref{fig:computer-results} which shows that the numerical integration agrees perfectly with the analytical results in the main text.


\begin{thebibliography}{10}
	
	\bibitem{funk2010modelling}
	Funk, S., Salathé, M., and Jansen, V. A.~A. (2010)
	Modelling the influence of human behaviour on the spread of infectious
	diseases: a review.
	{\em Journal of The Royal Society Interface,} {\bf 7}(50), 1247--1256.
	
	\bibitem{verelst2016behavioural}
	Verelst, F., Willem, L., and Beutels, P. (2016)
	Behavioural change models for infectious disease transmission: a systematic
	review (2010--2015).
	{\em Journal of The Royal Society Interface,} {\bf 13}(125).
	
	\bibitem{weston2018infection}
	Weston, D., Hauck, K., and Aml{\^o}t, R. (2018)
	Infection prevention behaviour and infectious disease modelling: a review of
	the literature and recommendations for the future.
	{\em BMC public health,} {\bf 18}(1), 336.
	
	\bibitem{reluga2010}
	Reluga, T. (2010)
	Game theory of social distancing in response to an epidemic.
	{\em PLoS Comput Biol.,} {\bf 6}(5).
	
	\bibitem{chen2011public}
	Chen, F., Jiang, M., Rabidoux, S., and Robinson, S. (2011)
	Public avoidance and epidemics: insights from an economic model.
	{\em Journal of theoretical biology,} {\bf 278}(1), 107--119.
	
	\bibitem{rhodes1997epidemic}
	Rhodes, C. and Anderson, R. (1997)
	Epidemic thresholds and vaccination in a lattice model of disease spread.
	{\em Theoretical Population Biology,} {\bf 52}(2), 101--118.
	
	\bibitem{zhao2015}
	Zhao, S., Wu, J., and Ben-Arieh, D. (2015)
	Modeling infection spread and behavioral change using spatial games.
	{\em Health Systems,} {\bf 4}(1), 41--53.
	
	\bibitem{XIA2019185}
	Xia, C., Wang, Z., Zheng, C., Guo, Q., Shi, Y., Dehmer, M., and Chen, Z. (2019)
	A new coupled disease-awareness spreading model with mass media on multiplex
	networks.
	{\em Information Sciences,} {\bf 471}, 185 -- 200.
	
	\bibitem{gomez2010discrete}
	G{\'o}mez, S., Arenas, A., Borge-Holthoefer, J., Meloni, S., and Moreno, Y.
	(2010)
	Discrete-time Markov chain approach to contact-based disease spreading in
	complex networks.
	{\em EPL (Europhysics Letters),} {\bf 89}(3), 38009.
	
	\bibitem{kan2017effects}
	Kan, J.-Q. and Zhang, H.-F. (2017)
	Effects of awareness diffusion and self-initiated awareness behavior on
	epidemic spreading-an approach based on multiplex networks.
	{\em Communications in Nonlinear Science and Numerical Simulation,} {\bf 44},
	193--203.
	
	\bibitem{hota2019}
	{Hota}, A.~R. and {Sundaram}, S. (2019)
	Game-Theoretic Vaccination Against Networked SIS Epidemics and Impacts of Human
	Decision-Making.
	{\em IEEE Transactions on Control of Network Systems,} {\bf 6}(4), 1461--1472.
	
	\bibitem{zhang2014effects}
	Zhang, H.-F., Wu, Z.-X., Tang, M., and Lai, Y.-C. (2014)
	Effects of behavioral response and vaccination policy on epidemic spreading-an
	approach based on evolutionary-game dynamics.
	{\em Scientific reports,} {\bf 4}, 5666.
	
	\bibitem{schimit2011vaccination}
	Schimit, P. and Monteiro, L. (2011)
	A vaccination game based on public health actions and personal decisions.
	{\em Ecological Modelling,} {\bf 222}(9), 1651--1655.
	
	\bibitem{trajanovski2015decentralized}
	Trajanovski, S., Hayel, Y., Altman, E., Wang, H., and Van~Mieghem, P. (2015)
	Decentralized protection strategies against SIS epidemics in networks.
	{\em IEEE Transactions on Control of Network Systems,} {\bf 2}(4), 406--419.
	
	\bibitem{liu2012impact}
	Liu, X.-T., Wu, Z.-X., and Zhang, L. (2012)
	Impact of committed individuals on vaccination behavior.
	{\em Physical Review E,} {\bf 86}(5), 051132.
	
	\bibitem{kabir2019}
	Kabir, K.~A. and Tanimoto, J. (2019)
	Dynamical behaviors for vaccination can suppress infectious disease--A game
	theoretical approach.
	{\em Chaos, Solitons \& Fractals,} {\bf 123}, 229--239.
	
	\bibitem{kuga2019vaccinate}
	Kuga, K., Tanimoto, J., and Jusup, M. (2019)
	To vaccinate or not to vaccinate: A comprehensive study of
	vaccination-subsidizing policies with multi-agent simulations and mean-field
	modeling.
	{\em Journal of theoretical biology,} {\bf 469}, 107--126.
	
	\bibitem{fu2011imitation}
	Fu, F., Rosenbloom, D.~I., Wang, L., and Nowak, M.~A. (2011)
	Imitation dynamics of vaccination behaviour on social networks.
	{\em Proceedings of the Royal Society B: Biological Sciences,} {\bf 278}(1702),
	42--49.
	
	\bibitem{hayashi2016effects}
	Hayashi, M.~A. and Eisenberg, M.~C. (2016)
	Effects of adaptive protective behavior on the dynamics of sexually transmitted
	infections.
	{\em Journal of theoretical biology,} {\bf 388}, 119--130.
	
	\bibitem{bhattacharyya2011wait}
	Bhattacharyya, S. and Bauch, C.~T. (2011)
	“Wait and see” vaccinating behaviour during a pandemic: A game theoretic
	analysis.
	{\em Vaccine,} {\bf 29}(33), 5519--5525.
	
	\bibitem{bauch2012evolutionary}
	Bauch, C.~T. and Bhattacharyya, S. (2012)
	Evolutionary game theory and social learning can determine how vaccine scares
	unfold.
	{\em PLoS computational biology,} {\bf 8}(4).
	
	\bibitem{bauch2005imitation}
	Bauch, C.~T. (2005)
	Imitation dynamics predict vaccinating behaviour.
	{\em Proceedings of the Royal Society B: Biological Sciences,} {\bf 272}(1573),
	1669--1675.
	
	\bibitem{li2017provisioning}
	Li, J., Lindberg, D.~V., Smith, R.~A., and Reluga, T.~C. (2017)
	Provisioning of public health can be designed to anticipate public policy
	responses.
	{\em Bulletin of mathematical biology,} {\bf 79}(1), 163--190.
	
	\bibitem{wang2013impact}
	Wang, Y., Cao, J., Jin, Z., Zhang, H., and Sun, G.-Q. (2013)
	Impact of media coverage on epidemic spreading in complex networks.
	{\em Physica A: Statistical Mechanics and its Applications,} {\bf 392}(23),
	5824--5835.
	
	\bibitem{wang2020effect}
	Wang, M., Pan, Q., and He, M. (2020)
	The effect of individual attitude on cooperation in social dilemma.
	{\em Physica A: Statistical Mechanics and its Applications,} p. 124424.
	
	\bibitem{jnawali2016emergence}
	Jnawali, K., Morsky, B., Poore, K., and Bauch, C.~T. (2016)
	Emergence and spread of drug resistant influenza: a two-population game
	theoretical model.
	{\em Infectious Disease Modelling,} {\bf 1}(1), 40--51.
	
	\bibitem{wu2011imperfect}
	Wu, B., Fu, F., and Wang, L. (2011)
	Imperfect vaccine aggravates the long-standing dilemma of voluntary
	vaccination.
	{\em PloS one,} {\bf 6}(6).
	
	\bibitem{cai2014effect}
	Cai, C.-R., Wu, Z.-X., and Guan, J.-Y. (2014)
	Effect of vaccination strategies on the dynamic behavior of epidemic spreading
	and vaccine coverage.
	{\em Chaos, Solitons \& Fractals,} {\bf 62}, 36--43.
	
	\bibitem{who}
	Chantal~Blouin, Nick~Drager, R.~S. (2006)
	International Trade in Health Services and the GATS: current issues and
	debates,
	The World Bank, .
	
	\bibitem{ebola}
	Richardson, E.~T., Barrie, M.~B., Nutt, C.~T., Kelly, J.~D., Frankfurter, R.,
	Fallah, M.~P., and Farmer, P.~E. (2017)
	The Ebola suspect's dilemma.
	{\em The Lancet Global Health,} {\bf 5}(3), e254--e256.
	
	\bibitem{asiatimes}
	Kaushik, P. (2020)
	Covid-19 and the Prisoner's Dilemma. \textit{Asia Times}.
	
	\bibitem{stubborn}
	Fukuda, E. and Tanimoto, J. (2015)
	Impact of Stubborn Individuals on a Spread of Infectious Disease under
	Voluntary Vaccination Policy.
	{\em Proceedings of the 18th Asia Pacific Symposium on Intelligent and
		Evolutionary Systems, Volume 1. Proceedings in Adaptation, Learning and
		Optimization,} {\bf 1}.
	
	\bibitem{nowak}
	Nowak, M.~A. (2006)
	Evolutionary dynamics,
	The Belknap Press of Harvard University Press, Cambridge, MA,
	Exploring the equations of life.
	
	\bibitem{axelrod}
	Axelrod, R. and Hamilton, W.~D. (1981)
	The evolution of cooperation.
	{\em Science,} {\bf 211}(4489), 1390--1396.
	
	\bibitem{axelrod2}
	Axelrod, R. (2012)
	Launching ``{T}he evolution of cooperation''.
	{\em J. Theoret. Biol.,} {\bf 299}, 21--24.
	
	\bibitem{cooperate1}
	Nee, S. (1989)
	Does {H}amilton's rule describe the evolution of reciprocal altruism?.
	{\em J. Theoret. Biol.,} {\bf 141}(1), 81--91.
	
	\bibitem{cooperate2}
	Sakai, S. (1987)
	Emergence of rational cooperation in the repeated prisoner's dilemma.
	{\em Bull. Koshien Univ. B,} (15), 35--40 (1988).
	
	\bibitem{cooperate3}
	Nowak, M.~A. (2006)
	Evolutionary dynamics of cooperation.
	In \emph{International {C}ongress of {M}athematicians. {V}ol. {III}} pp.
	1523--1540 Eur. Math. Soc., Z\"{u}rich.
	
	\bibitem{cooperate4}
	Rand, D.~G., Dreber, A., Ellingsen, T., Fudenberg, D., and Nowak, M.~A. (2009)
	Positive interactions promote public cooperation.
	{\em Science,} {\bf 325}(5945), 1272--1275.
	
	\bibitem{cooperate5}
	Nowak, M.~A. (2012)
	Evolving cooperation.
	{\em J. Theoret. Biol.,} {\bf 299}, 1--8.
	
	\bibitem{martin}
	Taylor, C. \&~Martin, A.~N. (2007)
	Transforming the dilemma.
	{\em Evolution,} {\bf 61}(10), 2281--2292.
	
	\bibitem{ore}
	Bravetti, A. \&~Padilla, P. (2018)
	An optimal strategy to solve the Prisoner's Dilemma.
	{\em Nature Scientific Reports,} {\bf 8}(1948).
	
	\bibitem{poletti2009}
	Poletti, P., Caprile, B., Ajelli, M., Pugliese, A., and Merler, S. (2009)
	Spontaneous behavioural changes in response to epidemics.
	{\em Journal of theoretical biology,} {\bf 260}(1), 31--40.
	
	\bibitem{lin2020conceptual}
	Lin, Q., Zhao, S., Gao, D., Lou, Y., Yang, S., Musa, S.~S., Wang, M.~H., Cai,
	Y., Wang, W., Yang, L., et al. (2020)
	A conceptual model for the coronavirus disease 2019 (COVID-19) outbreak in
	Wuhan, China with individual reaction and governmental action.
	{\em International journal of infectious diseases,} {\bf 93}, 211--216.
	
	\bibitem{hiv}
	Romer, D., Sznitman, S., Salazar, L.~F., Vanable, P.~A., Carey, M.~P.,
	Hennessy, M., Brown, L.~K., Valois, R.~F., Stanton, B.~F., Fortune, T., and
	Juzang, I. (2009)
	Mass Media as an HIV-Prevention Strategy: Using Culturally Sensitive Messages
	to Reduce HIV-Associated Sexual Behavior of At-Risk African American Youth.
	{\em American Journal of Public Health,} {\bf 12}(99), 2150--2159.
	
	\bibitem{mice}
	Barthold, S.W. \&~Smith, A. (1989)
	Duration of challenge immunity to coronavirus JHM in mice.
	{\em Archives of Virology,} (107), 171--177.
	
	\bibitem{vaccine}
	Barrett, Alan D.T. \&~Stanberry, L.~R. (2009)
	Vaccines for biodefense and emerging and neglected diseases,
	Academic Press.
	
\end{thebibliography}
\end{document}